\documentclass[11pt,conference,a4paper]{IEEEtran}

\usepackage[dvipsnames]{xcolor}
\usepackage{amsmath,graphicx}
\usepackage{amsfonts}
\usepackage{amssymb}
\usepackage{tikz}
\usetikzlibrary{arrows, decorations.markings,chains}
\usepackage{physics}
\usepackage{comment}
\usepackage{algorithm}
\usepackage{algorithmic}
\usepackage{array}
\usepackage{pgfplots}
\usepackage{color}

\usepackage{authblk}

\tikzstyle{vecArrow} = [thick, decoration={markings,mark=at position
   1 with {\arrow[semithick]{open triangle 60}}},
   double distance=1.4pt, shorten >= 5.5pt,
   preaction = {decorate},
   postaction = {draw,line width=1.4pt, white,shorten >= 4.5pt}]
\tikzstyle{innerWhite} = [semithick, white,line width=1.4pt, shorten >= 4.5pt]
\newcommand{\bs}[1]{\boldsymbol{#1}}

\begin{document}
\title{PSK Precoding in Multi-User MISO Systems}
%
\author[1]{Andreas~Noll}
\author[1]{Hela~Jedda}
\author[1,2]{Josef~A.~Nossek}
\affil[1]{Technical University of Munich, 80290 Munich, Germany}
\affil[2]{Federal University of Cear\'a, Fortaleza, Brazil}
\affil[ ] {Email: \{andreas.noll, hela.jedda, josef.a.nossek\}@tum.de}
%
%
%

%
\maketitle
\begin{abstract}
We consider the downlink scenario of multi-user multiple-input-single-output (MU-MISO) communication systems with constant envelope (CE) signals emitted from each antenna. This results in energy efficient power amplifiers (PAs). We propose a holistic CE precoding scheme based on the symbol-wise minimum squared error (SMSE) criterion. Additionally, we analyze the distortions introduced by low-resolution quantization to PSK for higher energy efficiency reasons. We present three solution algorithms and examine their performance to decide for the best pick for different quantization resolutions. Our results show that good performance can be achieved with minimal loss compared to an ideal unquantized case. Finally, we analyze and discuss the results and consider the overall complexity of the precoder as well as implementation issues.
\end{abstract}
\section{Introduction}
\label{sec:intro}
The ever increasing demand for higher data rates in mobile communications poses significant challenges for research. It is expected that the network capacity is increased 1000-fold and the number of connected devices 10-100-fold compared to 4G networks \cite{Osseiran2014}. 

This leads to higher energy consumption. Most of the energy is consumed by the base stations (BSs) \cite{RFWC}. Typically, the RF PA accounts for more than half of the energy consumption in a BS \cite{Blume2010, Chen2010}. 

The highest energy efficiency is achieved when the PA is operated in the saturation region. However, operation in that region implies high nonlinear distortions that are introduced to the signals. In the literature there exist several techniques for PA efficiency enhancement and nonlinear distortions minimization such as envelope
elimination and restoration \cite{FabbroKayal2010} and envelope tracking 
(ET) \cite{FabbroKayal2010}.

Another approach to have energy efficient PA is the CE modulation scheme at the PA input. Hence, the amplitude does not bear any information and the PA can operate in the saturation region with highest energy efficiency and linearity is not required. In \cite{Larsson2013} a precoding technique with continuous valued CE signals is introduced. This method aims at minimizing the multi-user interference (MUI), whereas the authors in \cite{AmadoriMasouros2016} have shown that constructive MUI is beneficial to improve the performance. In our contribution, we consider a different problem formulation based on minimizing the squared error between a scaled version of the desired vector and the noiseless receive vector.

Another important measure to achieve more energy efficient systems is the usage of low-resolution digital-to-analog converters (DACs). The CE signals have to be then quantized to PSK constellation. To the best of our knowledge, only 1-bit quantization has been considered so far. The contribution in \cite{Mezghani2009} is the first work that addressed the precoding task with 1-bit quantization at the transmitter. The authors in \cite{Usman2016} introduced another linear precoder that could slightly improve the system performance. Theoretical analysis on the achievable rate in systems with 1-bit transmitters were introduced in \cite{Kakkavas2016, Saxena2016, Yongzhi2016}. The first nonlinear precoding technique in this topic was presented in \cite{JeddaSAM2016}. The authors presented a symbol-wise precoding technique based on the so called minimum bit error ratio (MBER) criterion and made use of the infinity norm to relax the 1-bit constraint. In \cite{Jacobsson_Studer2016_1} the authors present another symbol-wise precoder based on the minimum mean square error (MMSE) and extended it to higher modulation scheme in \cite{Jacobsson_Studer2016_2}. In our contribution we generalize the scenario to have PSK signals instead of only QPSK signals at the transmitter. The optimization criterion is the symbol-wise MSE (SMSE).
In this work we consider a downlink massive MU-MISO system, since the large scale of transmit antennas enhances the energy efficiency significantly \cite{Rusek2013}. We develop a symbol-wise precoding scheme that provides good performance while being energy efficient. We investigate the effects of quantized CE signals on the performance. To achieve that we analyze different solution approaches and compare them with respect to performance and efficiency.

This paper is organized as follows: in Section \ref{sec:sysmodel} we present the system model. In Section \ref{sec:precoder} we formulate the optimization problem to design the precoder and introduce the solving algorithms. In Sections \ref{sec:results}, \ref{sec:discussion} and \ref{sec:conclusion} we show the simulation results, discuss these results and the complexity of the proposed precoding scheme and summarize this work.

\textbf{Notation}: Bold letters indicate vectors and matrices, non-bold letters express scalars. The operators $(.)^{*}$, $(.)^{\rm T}$ and $(.)^{\rm H}$ stand for complex conjugation, transposition and Hermitian transposition, respectively. The $n \times n$ identity (zeros) matrix is denoted by $\bs{I}_{n}$ ($\bs{0}_{n}$).

\section{System Model}
\label{sec:sysmodel}

\begin{figure}[h!]
	\centering
	\begin{tikzpicture}[thick, scale=0.8]
	
  	\draw[vecArrow] (-1,0.5) to (0,0.5);
  	\draw[vecArrow] (1,0.5) to (2,0.5);
	\draw[vecArrow] (3,0.5) to (4,0.5);
	\draw[vecArrow] (5,0.5) to (5.6,0.5);
	\draw[vecArrow] (6,0.5) to (7,0.5);
	\draw[vecArrow] (8,0.5) to (9,0.5);
	\draw[vecArrow] (5.8,-0.5) to (5.8,0.4);
	
	\draw (5.8,0.5) circle (0.2);
	\draw (5.8,0.5) node[] {\Large$+$};
	
	\draw (0,0) rectangle (1,1);
	\draw (2,0) rectangle (3,1);
	\draw (4,0) rectangle (5,1);
	\draw (7,0) rectangle (8,1);
	
	\draw (0.5,0.5) node[] {$\mathcal{M}$};
	\draw (2.5,0.5) node[] {$\mathcal{Q}$};
	\draw (4.5,0.5) node[] {$\bs{H}$};
	\draw (7.5,0.5) node[] {$\mathcal{D}$};
	
	\draw (-0.5,1) node[] {$\bs{s}$};
	\draw (1.5,1) node[] {$\bs{x}$};
	\draw (3.5,1) node[] {$\bs{x}_\mathcal{Q}$};
	\draw (6.5,1) node[] {$\bs{r}$};
	\draw (8.5,1) node[] {$\bs{\hat s}$};
	\draw (5.8,-0.5) node[below] {$\bs{\eta}$};
	
	\draw (-0.5,-0.1) node[] {$\mathbb{C}^M$};
	\draw (1.5,-0.1) node[] {$\mathbb{C}^N$};
	\draw (3.5,-0.1) node[] {$\mathbb{C}^N$};
	\draw (6.5,-0.1) node[] {$\mathbb{C}^M$};
	\draw (8.5,-0.1) node[] {$\mathbb{C}^M$};

\end{tikzpicture}
	\caption{System model of the MU-MISO case.}
	\label{fig:sysmodel}
\end{figure}
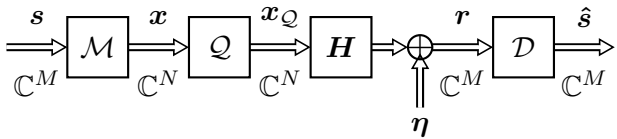

We consider the downlink MU-MISO system given in Fig. \ref{fig:sysmodel}. The BS has $N$ antennas and serves $M$ users with a single antenna each where $N \gg M$. The vector $\bs{s} \in \mathbb{C}^M$ contains the symbols for each user, where each symbol is drawn from the QPSK constellation. We assume that $\rm{E}[\bs{ss}^{\rm H}] = \sigma_s^2\bs{I}_M$. With a look-up-table (LUT) in $\mathcal{M}$ each input vector $\bs{s}$ is mapped to the vector $\bs{x}$ that fulfills the CE property, so we get
\begin{equation}
\bs{x} = \left[e^{j\phi_1},e^{j\phi_2},\dots,e^{j\phi_N}\right]^T = e^{j\bs{\phi}}. \label{eq:xCE}
\end{equation}
We introduce the quantization block $\mathcal{Q}$, to model the finite resolution of the DACs at the transmitter. The vector $\bs{x}$ is quantized by $\mathcal{Q}$ to $2^B$-PSK symbols, where $B$ denotes the quantizer resolution. We get 
\begin{equation}
\bs{x}_\mathcal{Q} = \mathcal{Q}(\bs{x}) = e^{j\mathcal{Q}_B(\bs{\phi})}.
\end{equation} The decoded signal is $\bs{\hat s} = \mathcal{D}\left(\sqrt{\frac{E_{tx}}{N}}\bs{H}\bs{x}_\mathcal{Q}+\bs{\eta}\right)$, where $\mathcal{D}$ is the decision operation of the QPSK constellation, $E_{tx}$ is the transmit energy, $\bs{H}$ represents the channel matrix and $\bs{\eta} \sim \mathcal{CN}(\bs{0}_M,\bs{I}_M)$ is the noise vector. We assume an i.i.d. Rayleigh-fading channel with zero mean and unit variance for each coefficient.

\section{CE Precoding Scheme}
\label{sec:precoder}

\subsection{Problem Formulation}

To determine the LUT for $\mathcal{M}$, the SMSE criterion is applied. The optimal CE transmit vector $\bs{x}$ is calculated as a function of $\bs{s}$ such that the squared error between a scaled version of the symbol and the noiseless channel output is minimized under some constraint
\begin{equation}
\begin{array}{rrclcl}
\displaystyle \min_{\bs{x}} & \multicolumn{3}{l}{\Vert\alpha\bs{s}-\bs{Hx}\Vert_2^2} \\
\textrm{s.t.} & x_i \in \mathcal{S} ,& \forall i = 1,\dots,N. &
\end{array}\label{eq:OpProblem}
\end{equation}
The factor $\alpha \in \mathbb{R}^+$ is introduced to utilize the channel gain more efficiently \cite{prabhu2015}. The set $\mathcal{S}$ denotes the constraint set for every $x_i$. For each $\bs{s}$ the respective solution $\bs{x}$ to the problem (\ref{eq:OpProblem}) is written into the LUT. The LUT has a dimension of $N \cross 4^M$, since we have $4^M$ distinct input vectors $\bs{s}$. In the following we present the choices for the constraint set $\mathcal{S}$ and the used algorithms we want to analyze.

\subsection{Constraint Set $\mathcal{S}$}

We have two choices for our constraint set. The first is to only allow for CE values without relaxation and the other is to relax the set and take the quantization into account.

\subsubsection{CE Constraint (CEC)}

The CEC can be written as
\begin{equation}
|x_i|^2 = 1, \forall i = 1,\dots,N.
\end{equation}
The solution to (\ref{eq:OpProblem}) is of the form given in (\ref{eq:xCE}).
Inserting (\ref{eq:xCE}) into (\ref{eq:OpProblem}) results in
\begin{equation}
\min_{\bs{\phi}}\Vert\alpha\bs{s}-\sum_{n=1}^N \bs{h}_ne^{j\phi_n}\Vert_2^2, \label{eq:OPP}
\end{equation}
where $\bs{h}_n$ is the $n$-th column of $\bs{H}$. Note that (\ref{eq:OPP}) looks similar to the optimization problem in \cite{Larsson2013}. However, we additionally introduce the scaling factor $\alpha$. We then solve the problem using either Algorithm \ref{alg:GDM} or Algorithm \ref{alg:QGDM}.

\subsubsection{Relaxed Polygon Constraint (RPC)}

For this case we take into consideration that the transmit vector $\bs{x}$ is quantized to $\bs{x}_\mathcal{Q}$. We define $\mathcal{S}$ as the filled polygon built by the points of the $2^B$-PSK constellation. Hence, we allow that each entry of $\bs{x}$ be within that polygon. After the solution algorithm is run, the resulting entries in $\bs{x}$ are normalized to CE and quantized to $2^B$-PSK. If there is no quantization present, the polygon  becomes the unit circle. To solve the problem with the RPC we use Algorithm \ref{alg:GPM}.

\subsection{Solution Algorithms}

\subsubsection{Gradient Descent Method (GDM)}

This method is used to solve the problem in (\ref{eq:OPP}) with the CEC. The GDM is suitable, because even local minima are close to optimal \cite{Larsson2013}. By defining the objective function as 
\begin{equation}
g(\bs{\phi},\bs{s}) := \|\alpha\bs{s}-\sum_{n=1}^N \bs{h}_ne^{j\phi_n}\|^2_2,
\end{equation}
the gradient is expressed in closed form as
\begin{equation}
\frac{\partial g(\bs{\phi},\bs{s})}{\partial\bs{\phi}} = \left[\frac{\partial g(\bs{\phi},\bs{s})}{\partial\phi_1},\frac{\partial g(\bs{\phi},\bs{s})}{\partial\phi_2},\dots,\frac{\partial g(\bs{\phi},\bs{s})}{\partial\phi_N}\right]^T
\end{equation}
with the partial derivatives
\begin{equation}
\frac{\partial g(\bs{\phi},\bs{s})}{\partial\phi_n}\! = \!\!-2\Im{\!e^{-j\phi_n}\bs{h}_n^H\!\left(\alpha\bs{s}-\sum_{m=1}^N \bs{h}_me^{j\phi_m}\!\right)}.
\end{equation}
The GDM is given in Algorithm \ref{alg:GDM}.

\begin{algorithm}
\caption{Gradient descent method}
\label{alg:GDM}
\begin{algorithmic}
\REQUIRE Symbol vector $\bs{s}$, Number of antennas $N$, Number of users $M$, Channel matrix $\bs{H}$
\ENSURE Transmit vector $\bs{x}$
\STATE Step size $\mu = \mu_0$, Tolerable error $\epsilon$, Scaling factor $\alpha$, $\bs{\phi}^{(0)} = \bs{0}$, $n = 0$
\WHILE{$err > \epsilon$} \STATE{$\bs{\phi}^{(n+1)} = \bs{\phi}^{(n)}-\mu\frac{\partial g(\bs{\phi}^{(n)},\bs{s})}{\partial\bs{\phi}^{(n)}}$}
\STATE $err = \|\bs{\phi}^{(n+1)}-\bs{\phi}^{(n)}\|$
	\IF{$g(\bs{\phi}^{(n+1)},\bs{s}) > g(\bs{\phi}^{(n)},\bs{s})$}
 	\STATE{$\mu = \mu/2$}
	\STATE{$\bs{\phi}^{(n+1)} = \bs{\phi}^{(n)}$}
	\ENDIF
\STATE $n = n+1$
\ENDWHILE
\STATE $\bs{x} = e^{j\bs{\phi}^{(n)}}$
\end{algorithmic}
\end{algorithm}

\subsubsection{Quantized Gradient Descent Method (QGDM)}

The QGDM is also used to solve (\ref{eq:OPP}) with the CEC. It is given in Algorithm \ref{alg:QGDM}. It involves the quantization operation $\mathcal{Q}_B$ after every gradient step. Thus, in each iteration step we have $2^B$-PSK symbols in the transmit vector. The objective function is the same $g(\bs{\phi},\bs{s})$. 

If a step is successful, i.e. the value of $g(\bs{\phi},\bs{s})$ is reduced, the step size $\mu$ is reset again to the starting value $\mu_0$. This is because the gradient together with the step size is quantized. We found that with this reset the performance can be improved. 

\begin{algorithm}
\caption{Quantized gradient descent method}
\label{alg:QGDM}
\begin{algorithmic}
\REQUIRE Symbol vector $\bs{s}$, Number of antennas $N$, Number of users $M$, Channel matrix $\bs{H}$, Quantization resolution $B$
\ENSURE Transmit vector $\bs{x}$
\STATE Step size $\mu = \mu_0$, Tolerable error $\epsilon$, Scaling factor $\alpha$, $\bs{\phi}^{(0)} = \bs{0}$, $n = 0$
\WHILE{$err > \epsilon$} 
\STATE{$\bs{\phi}^{(n+1)} = \mathcal{Q}_B\left(\bs{\phi}^{(n)}-\mu\frac{\partial g(\bs{\phi}^{(n)},\bs{s})}{\partial\bs{\phi}^{(n)}}\right)$}
\STATE $err = \|\bs{\phi}^{(n+1)}-\bs{\phi}^{(n)}\|$
	\IF{$g(\bs{\phi}^{(n+1)},\bs{s}) > g(\bs{\phi}^{(n)},\bs{s})$}
 	\STATE{$\mu = \mu/2$}
	\STATE{$\bs{\phi}^{(n+1)} = \bs{\phi}^{(n)}$}
	\ELSE
	\STATE{$\mu = \mu_0$}
	\ENDIF
\STATE $n = n+1$
\ENDWHILE
\STATE $\bs{x} = e^{j\bs{\phi}^{(n)}}$
\end{algorithmic}
\end{algorithm}

\subsubsection{Gradient Projection Method (GPM)}

The GPM is used to solve (\ref{eq:OpProblem}) with the RPC. It operates directly on the vector $\bs{x}$ and involves a projection in every step of entries that fall outside the polygon back onto the boundary of it. This projection operation is denoted by $\mathcal{P}_B$ as the shape of the polygon depends on $B$. 

The objective function is defined as
\begin{equation}
f(\bs{x},\bs{s}) := \Vert\alpha\bs{s}-\bs{Hx}\Vert_2^2
\end{equation}
and the gradient can be expressed in closed form as
\begin{equation}
\frac{\partial f(\bs{x},\bs{s})}{\partial\bs{x}} = -\alpha\bs{H}^T\bs{s}^*+\bs{H}^T\bs{H}^*\bs{x}^*.
\end{equation}
The GPM is given in Algorithm \ref{alg:GPM}.

\begin{algorithm}
\caption{Gradient projection method}
\label{alg:GPM}
\begin{algorithmic}
\REQUIRE Symbol vector $\bs{s}$, Number of antennas $N$, Number of users $M$, Channel matrix $\bs{H}$, Quantization resolution $B$
\ENSURE Transmit vector $\bs{x}$
\STATE Step size $\mu = \mu_0$, Tolerable error $\epsilon$, Scaling factor $\alpha$, $\bs{x}^{(0)} = \mathcal{P}_B\{\bs{1}\}$, $n = 0$
\WHILE{$err > \epsilon$} 
\STATE{$\bs{x}^{(n+1)} = \mathcal{P}_B\left\{\bs{x}^{(n)}-\mu\left(\frac{\partial f(\bs{x}^{(n)},\bs{s})}{\partial\bs{x}^{(n)}}\right)^*\right\}$}
\STATE $err = \|\bs{x}^{(n+1)}-\bs{x}^{(n)}\|$
	\IF{$f(\bs{x}^{(n+1)},\bs{s}) > f(\bs{x}^{(n)},\bs{s})$}
 	\STATE{$\mu = \mu/2$}
	\STATE{$\bs{x}^{(n+1)} = \bs{x}^{(n)}$}
	\ENDIF
\STATE $n = n+1$
\ENDWHILE
\STATE $\bs{x} = \bs{x}^{(n)}$
\end{algorithmic}
\end{algorithm}

\subsection{Approximation of $\alpha$}
In this work, we choose $\alpha$ to be equal to the expectation value of the scaling factor in the case of a zero-forcing precoder \cite{Joham2005}
\begin{align}
\alpha &= \text{ E} \left\lbrace \sqrt{\frac{\sum_{i=1}^N |x_i|^2}{\trace\left( \left(\bs H \bs H^{\text{H}}\right)^{-1}\right)}} \right\rbrace.
\end{align}
According to (2.9) in \cite{Tulino_Verdu2004} and since $ \left(\bs H \bs H^{\text{H}}\right)^{-1}$ is a Wishart matrix with $N>M$, we get
\begin{align}
\text{ E} \left\lbrace \trace\left( \left(\bs H \bs H^{\text{H}}\right)^{-1}\right) \right\rbrace &= \frac{M}{N-M}.
\label{eq:wishart}
\end{align}
Considering the CEC and the result in (\ref{eq:wishart}) we get
\begin{align}
\alpha &= \sqrt{\frac{N \left( N-M\right)}{M}}.
\end{align}

\subsection{Existing Precoder}
\label{subsec:wfprecoder}

We compare our proposed method to the existing Wiener Filter (WF) precoder as the ideal case without quantization and without CE constraint. The WF precoder is linear and defined by
\begin{equation}
\bs{x} = \bs{P}_{WF}\bs{s} \:\: \text{   and   } \:\:  \bs{\hat s} = {f}_{WF}\bs{r}
\end{equation}
with
\begin{equation}
\begin{aligned}
\bs{P}_{WF} &= \frac{1}{f_{WF}}\left(\bs{H}^H\bs{H}+\frac{M\bs{I}_N}{E_{tx}}\right)^{-1}\bs{H}^H,\\
f_{WF} &= \sqrt{\frac{\sigma_s^2}{E_{tx}}\tr\left[\left(\bs{H}^H\bs{H}+\frac{M\bs{I}_N}{E_{tx}}\right)^{-2}\bs{H}^H\bs{H}\right]}.
\end{aligned}
\end{equation}

\section{Simulation Results}
\label{sec:results}

We simulate the proposed precoding scheme. All results are averaged over 500 channel realizations. The symbol energy is $\sigma_s^2=1$ and over each channel a total of $N_b = 10^4$ symbols are sent. We have a total of $N = 32$ antennas and $M = 4$ users. For the error tolerance we set $\epsilon = 10^{-2}$ for all three methods. In the GDM we set for the starting step size $\mu_0 = 0.25$, the QGDM has $\mu_0 = 0.5$ and the GPM starts with $\mu_0 = 1$. As the performance measure we use the uncoded BER. 

We observed that for 16-PSK the performance differs by less than 0.2 dB from the unquantized CE case. Thus, we show the results for $B \in \{2,3,\infty\}$ and for values of $B > 3$ performance is almost identical to $B = \infty$. 

We denote each precoder by its corresponding algorithm and constraint set, i.e. GDM CEC, QGDM CEC and GPM RPC.

\begin{figure}
	\centering
%
%
\definecolor{mycolor1}{rgb}{0.00000,1.00000,1.00000}%
\definecolor{mycolor2}{rgb}{1.00000,1.00000,0.00000}%
\begin{tikzpicture}

\begin{axis}[%
width=0.8\columnwidth,
height=5cm,
at={(1.011in,0.642in)},
scale only axis,
xmin=-10,
xmax=20,
xlabel={SNR (dB)},
xmajorgrids,
ymode=log,
ymin=1e-04,
ymax=1,
yminorticks=true,
ylabel={Uncoded BER},
ymajorgrids,
yminorgrids,
axis background/.style={fill=white},
legend style={legend cell align=left,align=left,draw=white!15!black}
]
\addplot [color=orange,solid,line width=1.0pt,mark=10-pointed star,smooth]
  table[row sep=crcr]{%
-10	0.37591295\\
-8	0.345329625\\
-6	0.308115725\\
-4	0.2640238\\
-2	0.21344535\\
0	0.158646175\\
2	0.104019075\\
4	0.056504875\\
6	0.022995125\\
8	0.00600085\\
10	0.00078195\\
12	3.24e-05\\
14	1.75e-07\\
16	0\\
18	0\\
20	0\\
};
\addlegendentry{M\&L CE};

\addplot [color=Plum,solid,line width=1.0pt,mark=o,smooth]
  table[row sep=crcr]{%
-10	0.220610225\\
-8	0.168986325\\
-6	0.117266275\\
-4	0.0711864\\
-2	0.03611755\\
0	0.014693575\\
2	0.00465685\\
4	0.001218325\\
6	0.0002898\\
8	7.26e-05\\
10	2.065e-05\\
12	7.7e-06\\
14	3.175e-06\\
16	1.525e-06\\
18	1.025e-06\\
20	4.5e-07\\
};
\addlegendentry{WF CE};

\addplot [color=YellowOrange,solid,line width=1.0pt,mark=square,smooth]
  table[row sep=crcr]{%
-10	0.22519025\\
-8	0.1715819\\
-6	0.117030575\\
-4	0.0680433249999999\\
-2	0.03136285\\
0	0.010372175\\
2	0.00216295\\
4	0.000262675\\
6	1.745e-05\\
8	7e-07\\
10	7.5e-08\\
12	0\\
14	0\\
16	0\\
18	0\\
20	0\\
};
\addlegendentry{GPM RPC};

\addplot [color=red,solid,line width=1.0pt,mark=triangle,smooth]
  table[row sep=crcr]{%
-10	0.225465375\\
-8	0.1718344\\
-6	0.117218125\\
-4	0.06812385\\
-2	0.031338\\
0	0.010321625\\
2	0.0021294\\
4	0.000252675\\
6	1.595e-05\\
8	6.5e-07\\
10	5e-08\\
12	0\\
14	0\\
16	0\\
18	0\\
20	0\\
};
\addlegendentry{GDM CEC};

\addplot [color=blue,solid,line width=1.0pt,mark=diamond,smooth]
  table[row sep=crcr]{%
-10	0.192202225\\
-8	0.13861645\\
-6	0.0871717249999999\\
-4	0.04475245\\
-2	0.016883825\\
0	0.003983875\\
2	0.000450375\\
4	1.80500000000001e-05\\
6	7.5e-08\\
8	0\\
10	0\\
12	0\\
14	0\\
16	0\\
18	0\\
20	0\\
};
\addlegendentry{WF unq.};

\end{axis}
\end{tikzpicture}%
	\caption{Uncoded BER over the SNR for different precoder designs for $N =32$ and $M=4$ and $B = \infty$.}
	\label{fig:BER1}
\end{figure}
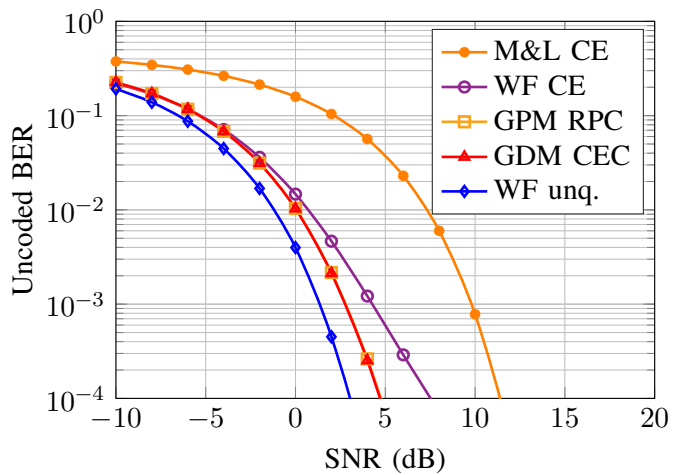

In a first simulation we omit the quantization block. Thus, the quantization resolution is $B = \infty$. We compare the GDM CEC precoder and the GPM RPC precoder to the WF precoder, the WF precoder with subsequent forcing of CE (WF CE) and the precoder proposed in \cite{Larsson2013} (M\&L CE). The results are given in Fig. \ref{fig:BER1}. The GDM CEC and GPM RPC precoders perform practically equally well. They clearly outperform the WF CE precoder and the M\&L CE precoder. At an uncoded BER of $10^{-3}$ the loss of the GDM CEC precoder to the ideal WF precoder is around 1.5 dB.

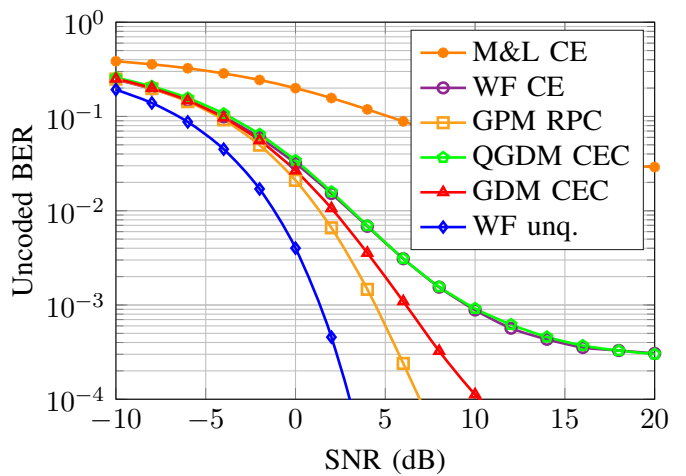
\begin{figure}
	\centering
%
%
\definecolor{mycolor1}{rgb}{0.00000,1.00000,1.00000}%
\definecolor{mycolor2}{rgb}{1.00000,1.00000,0.00000}%
\begin{tikzpicture}

\begin{axis}[%
width=0.8\columnwidth,
height=5cm,
at={(1.011in,0.642in)},
scale only axis,
xmin=-10,
xmax=20,
xlabel={SNR (dB)},
xmajorgrids,
ymode=log,
ymin=1e-04,
ymax=1,
yminorticks=true,
ylabel={Uncoded BER},
ymajorgrids,
yminorgrids,
axis background/.style={fill=white},
legend style={legend cell align=left,align=left,draw=white!15!black}
]
\addplot [color=orange,solid,line width=1.0pt,mark=10-pointed star,smooth]
  table[row sep=crcr]{%
-10	0.3852535625\\
-8	0.357477\\
-6	0.324523625\\
-4	0.2861636875\\
-2	0.243803875\\
0	0.199688\\
2	0.156783625\\
4	0.1187915\\
6	0.088751625\\
8	0.0668646875\\
10	0.0520455625\\
12	0.042458375\\
14	0.0365388125\\
16	0.0328045\\
18	0.0304760625\\
20	0.02903575\\
};
\addlegendentry{M\&L CE};

\addplot [color=Plum,solid,line width=1.0pt,mark=o,smooth]
  table[row sep=crcr]{%
-10	0.2455165\\
-8	0.196535875\\
-6	0.1462831875\\
-4	0.0988685624999999\\
-2	0.0597081875\\
0	0.0317830625\\
2	0.0152099375\\
4	0.006838375\\
6	0.003096125\\
8	0.0015444375\\
10	0.0008845625\\
12	0.0005664375\\
14	0.0004325\\
16	0.000354\\
18	0.0003288125\\
20	0.0003065625\\
};
\addlegendentry{WF CE};

\addplot [color=YellowOrange,solid,line width=1.0pt,mark=square,smooth]
  table[row sep=crcr]{%
-10	0.24914875\\
-8	0.197787125\\
-6	0.143695\\
-4	0.0921815\\
-2	0.0495904375\\
0	0.0210269375\\
2	0.006589875\\
4	0.001466125\\
6	0.0002398125\\
8	3.49375e-05\\
10	4.625e-06\\
12	6.25e-07\\
14	6.25e-08\\
16	0\\
18	0\\
20	0\\
};
\addlegendentry{GPM RPC};

\addplot [color=green,solid,line width=1.0pt,mark=pentagon,smooth]
  table[row sep=crcr]{%
-10	0.25801\\
-8	0.2087875\\
-6	0.15687725\\
-4	0.1068603125\\
-2	0.064639\\
0	0.0340340625\\
2	0.015881625\\
4	0.0069409375\\
6	0.003106375\\
8	0.001561625\\
10	0.0009185\\
12	0.0006199375\\
14	0.0004583125\\
16	0.0003711875\\
18	0.000328\\
20	0.0003013125\\
};
\addlegendentry{QGDM CEC};

\addplot [color=red,solid,line width=1.0pt,mark=triangle,smooth]
  table[row sep=crcr]{%
-10	0.2507934375\\
-8	0.2004334375\\
-6	0.1475985\\
-4	0.0973764375\\
-2	0.0557143125\\
0	0.02663275\\
2	0.0106020625\\
4	0.0035614375\\
6	0.0010919375\\
8	0.000324375\\
10	0.000112375\\
12	4.9625e-05\\
14	2.5625e-05\\
16	1.775e-05\\
18	1.375e-05\\
20	1.26875e-05\\
};
\addlegendentry{GDM CEC};

\addplot [color=blue,solid,line width=1.0pt,mark=diamond,smooth]
  table[row sep=crcr]{%
-10	0.192681\\
-8	0.138989\\
-6	0.087476625\\
-4	0.0449708125\\
-2	0.017018625\\
0	0.0040090625\\
2	0.000456\\
4	1.675e-05\\
6	6.25e-08\\
8	0\\
10	0\\
12	0\\
14	0\\
16	0\\
18	0\\
20	0\\
};
\addlegendentry{WF unq.};

\end{axis}
\end{tikzpicture}%
	\caption{Uncoded BER over the SNR for different precoder designs for $N =32$ and $M=4$ and $B = 2$.}
	\label{fig:BER2}
\end{figure}

Next, we set $B=2$ and have QPSK transmit signals. We now additionally compare the QGDM CEC precoder. The results are shown in Fig. \ref{fig:BER2}. Now the best precoder clearly is the GPM RPC. The QGDM CEC precoder is as good as the WF CE precoder. Here the loss of the GPM RPC precoder to the ideal WF precoder is around 3.2 dB at a BER of $10{-3}$.

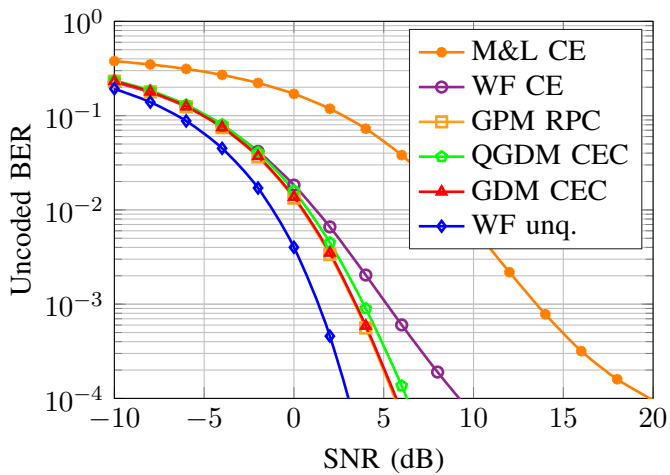
\begin{figure}
	\centering
%
%
\definecolor{mycolor1}{rgb}{0.00000,1.00000,1.00000}%
\definecolor{mycolor2}{rgb}{1.00000,1.00000,0.00000}%
\begin{tikzpicture}

\begin{axis}[%
width=0.8\columnwidth,
height=5cm,
at={(1.011in,0.642in)},
scale only axis,
xmin=-10,
xmax=20,
xlabel={SNR (dB)},
xmajorgrids,
ymode=log,
ymin=1e-04,
ymax=1,
yminorticks=true,
ylabel={Uncoded BER},
ymajorgrids,
yminorgrids,
axis background/.style={fill=white},
legend style={legend cell align=left,align=left,draw=white!15!black}
]
\addplot [color=orange,solid,line width=1.0pt,mark=10-pointed star,smooth]
  table[row sep=crcr]{%
-10	0.378875666666667\\
-8	0.348962333333333\\
-6	0.312932166666667\\
-4	0.270362166666667\\
-2	0.22214875\\
0	0.170174333333333\\
2	0.118366583333333\\
4	0.07263375\\
6	0.038097\\
8	0.0169010833333333\\
10	0.00638283333333333\\
12	0.00218433333333333\\
14	0.000781916666666667\\
16	0.000317666666666667\\
18	0.000159916666666667\\
20	9.59166666666667e-05\\
};
\addlegendentry{M\&L CE};

\addplot [color=Plum,solid,line width=1.0pt,mark=o,smooth]
  table[row sep=crcr]{%
-10	0.227184333333333\\
-8	0.176149666666667\\
-6	0.124640333333333\\
-4	0.0778835\\
-2	0.0416911666666667\\
0	0.0183166666666667\\
2	0.00660408333333334\\
4	0.00203675\\
6	0.00060225\\
8	0.000190416666666667\\
10	6.8e-05\\
12	3.40833333333333e-05\\
14	2.1e-05\\
16	1.45e-05\\
18	1.35833333333333e-05\\
20	1.20833333333333e-05\\
};
\addlegendentry{WF CE};

\addplot [color=YellowOrange,solid,line width=1.0pt,mark=square,smooth]
  table[row sep=crcr]{%
-10	0.2314565\\
-8	0.178401416666667\\
-6	0.124115833333333\\
-4	0.0745066666666666\\
-2	0.0364485\\
0	0.0133256666666667\\
2	0.003362\\
4	0.000559333333333334\\
6	6.575e-05\\
8	6.41666666666667e-06\\
10	5.83333333333333e-07\\
12	0\\
14	0\\
16	0\\
18	0\\
20	0\\
};
\addlegendentry{GPM RPC};

\addplot [color=green,solid,line width=1.0pt,mark=pentagon,smooth]
  table[row sep=crcr]{%
-10	0.2369205\\
-8	0.18446725\\
-6	0.13032025\\
-4	0.0801308333333334\\
-2	0.0408141666666667\\
0	0.0159585\\
2	0.00449766666666667\\
4	0.000899416666666667\\
6	0.000136166666666667\\
8	1.55833333333333e-05\\
10	2.41666666666667e-06\\
12	4.16666666666667e-07\\
14	0\\
16	0\\
18	0\\
20	0\\
};
\addlegendentry{QGDM CEC};

\addplot [color=red,solid,line width=1.0pt,mark=triangle,smooth]
  table[row sep=crcr]{%
-10	0.232245583333333\\
-8	0.179297583333333\\
-6	0.125136833333333\\
-4	0.0753214166666667\\
-2	0.0370936666666667\\
0	0.0137101666666667\\
2	0.00350208333333333\\
4	0.00058725\\
6	7.18333333333333e-05\\
8	6.58333333333334e-06\\
10	4.16666666666667e-07\\
12	0\\
14	0\\
16	0\\
18	0\\
20	0\\
};
\addlegendentry{GDM CEC};

\addplot [color=blue,solid,line width=1.0pt,mark=diamond,smooth]
  table[row sep=crcr]{%
-10	0.192684416666667\\
-8	0.139049916666667\\
-6	0.0876433333333333\\
-4	0.04499525\\
-2	0.0170628333333333\\
0	0.0040095\\
2	0.000458083333333333\\
4	1.775e-05\\
6	8.33333333333333e-08\\
8	0\\
10	0\\
12	0\\
14	0\\
16	0\\
18	0\\
20	0\\
};
\addlegendentry{WF unq.};

\end{axis}
\end{tikzpicture}%
	\caption{Uncoded BER over the SNR for different precoder designs for $N =32$ and $M=4$ and $B = 3$.}
	\label{fig:BER3}
\end{figure}

Finally, we simulate for $B = 3$ so  we have 8-PSK transmit signals. The results are shown in Fig. \ref{fig:BER3}. The GDM CEC precoder and the GPM RPC precoder perform equally well, closely followed by the QGDM CEC precoder. The loss of the GDM CEC or GPM RPC precoders to the ideal WF precoder is around 2.1 dB at a BER of $10^{-3}$.

We also simulated for $N = 60$ and $M = 6$ for comparison. The results were similar to the case with $N = 32$ and $M = 4$, so we discuss these results further. The relations in performance between the compared algorithms were equivalent.

\section{Discussion}
\label{sec:discussion}

In this section we analyze and discuss the results further. We will do so by comparing the speed of the algorithms and giving an outlook on the overall complexity of the precoding scheme.

\subsection{Algorithm Speed}
\label{subsec:parameter}

In practical applications the coherence time of the channel is limited. It is crucial that the algorithms converge fast and in consequence have a small number of iterations. 

First, we analyze for $B=\infty$. In table \ref{tab:iter1} we give the average number of iterations each algorithm requires. In table \ref{tab:skip1} we give the average number of times the step size is halved. In table \ref{tab:snr1} we give the SNR required to achieve a BER of $10^{-3}$. 

\begin{table}
	\centering
	\begin{tabular}{|c|c|}
	\hline Algorithm & Avg. nb. of iterations \\ \hline
	GDM & 39\\ \hline
	GPM & 46 \\ \hline
	\end{tabular}
	\caption{Average number of iterations of the two different algorithms and $B=\infty$.}
	\label{tab:iter1}
\end{table}

\begin{table}
	\centering
	\begin{tabular}{|c|c|}
	\hline Algorithm & Avg. nb. of halving \\ \hline
	GDM & 4 \\ \hline
	GPM & 5 \\ \hline
	\end{tabular}
	\caption{Average number of times the step size is halved for the two different algorithms and $B=\infty$.}
	\label{tab:skip1}
\end{table}

\begin{table}
	\centering
	\begin{tabular}{|c|c|}
	\hline Algorithm & SNR @ BER = $10^{-3}$ \\ \hline
	GDM & 2.71 dB \\ \hline
	GPM & 2.73 dB \\ \hline
	\end{tabular}
	\caption{Required SNR for a BER of $10^{-3}$ for the two different algorithms and $B=\infty$.}
	\label{tab:snr1}
\end{table}

The performance of the GDM and GPM is practically identical. On the other hand, the GDM is faster because the number of iterations is lower and in addition the performed operations are less complex. This is due to the high complexity of the projection operation $\mathcal{P}_B$. Thus, for the case of $B = \infty$, the GDM CEC precoder is the best choice.

Second, we analyze for $B = 2$. The respective results are in tables \ref{tab:iter2}, \ref{tab:skip2} and \ref{tab:snr2}.

\begin{table}
	\centering
	\begin{tabular}{|c|c|}
	\hline Algorithm & Avg. nb. of iterations \\ \hline
	GDM & 39\\ \hline
	QGDM & 14 \\ \hline
	GPM & 58 \\ \hline
	\end{tabular}
	\caption{Average number of iterations of the three different algorithms and $B=2$.}
	\label{tab:iter2}
\end{table}

\begin{table}
	\centering
	\begin{tabular}{|c|c|}
	\hline Algorithm & Avg. nb. of halving \\ \hline
	GDM & 4 \\ \hline
	QGDM & 10 \\ \hline
	GPM & 4 \\ \hline
	\end{tabular}
	\caption{Average number of times the step size is halved for the three different algorithms and $B=2$.}
	\label{tab:skip2}
\end{table}

\begin{table}
	\centering
	\begin{tabular}{|c|c|}
	\hline Algorithm & SNR @ BER = $10^{-3}$ \\ \hline
	GDM & 6.12 dB \\ \hline
	QGDM & 9.65 dB \\ \hline
	GPM & 4.46 dB \\ \hline
	\end{tabular}
	\caption{Required SNR for a BER of $10^{-3}$ for the three different algorithms and $B=2$.}
	\label{tab:snr2}
\end{table}

The best BER performance is achieved with the GPM algorithm, but the number of iterations is high. The QGDM CEC precoder is faster, but its performance is very poor. Thus, in this case it depends on the specific hardware and the efficiency of the projcetion $\mathcal{P}_B$, whether the GPM or the GDM should be chosen.

Last, we analyze for $B = 3$. The respective results are in tables \ref{tab:iter}, \ref{tab:skip} and \ref{tab:snr}.

\begin{table}
	\centering
	\begin{tabular}{|c|c|}
	\hline Algorithm & Avg. nb. of iterations \\ \hline
	GDM & 39 \\ \hline
	QGDM & 22 \\ \hline
	GPM & 55 \\ \hline
	\end{tabular}
	\caption{Average number of iterations of the three different algorithms and $B=3$.}
	\label{tab:iter}
\end{table}

\begin{table}
	\centering
	\begin{tabular}{|c|c|}
	\hline Algorithm & Avg. nb. of halving \\ \hline
	GDM & 4 \\ \hline
	QGDM & 17 \\ \hline
	GPM & 6 \\ \hline
	\end{tabular}
	\caption{Average number of times the step size is halved for the three different algorithms and $B=3$.}
	\label{tab:skip}
\end{table}

\begin{table}
	\centering
	\begin{tabular}{|c|c|}
	\hline Algorithm & SNR @ BER = $10^{-3}$ \\ \hline
	GDM & 3.39 dB \\ \hline
	QGDM & 3.86 dB \\ \hline
	GPM & 3.37 dB \\ \hline
	\end{tabular}
	\caption{Required SNR for a BER of $10^{-3}$ for the three different algorithms and $B=3$.}
	\label{tab:snr}
\end{table}

The algorithms with the best performance with respect to BER are the GDM and the GPM. They perform practically equally well. Clearly the fastest algorithm is the QGDM and the difference in SNR compared to the GDM or GPM is less than 0.5 dB. These results suggest that the QGDM is a good compromise with relatively low complexity, fast implementation and good performance.

We found that the number of iterations is almost invariant of the starting step size $\mu_0$. This is due to the fact that the step size is not reduced many times in the algorithms. We picked the different starting step sizes for each algorithm, because the performance was the best with these choices. The changes in performance with different $\mu_0$ were negligible for the GDM and GPM, but around 0.3 dB for the QGDM, because the quantization is present.

\subsection{Outlook on integral Complexity}
\label{subsec:complexity}
In the literature known to us, for example \cite{Larsson2013}, complexity is considered per symbol vector $\bs{s}$. Both the known algorithms as well as our proposed algorithms have low complexity for one $\bs{s}$. However, it is necessary to also consider the overall complexity as the precoding algorithms have to be run for each $\bs{s}$.

In a brute force implementation we would run the algorithms for every possible vector $\bs{s}$. For QPSK we have $4^M$ possible $\bs{s}$ and thus we have exponential complexity. 

We can reduce the number of times the algorithms have to be executed. For QPSK constellation we can construct some symbol vectors as simply rotated versions of other vectors. For any given vector $\bs{s}^*$ we can find three other vectors $\bs{s}'$, $\bs
{s}''$ and $\bs{s}'''$ for which we have $\bs{s}' = e^{j\frac{\pi}{2}}\bs{s}^*$, $\bs{s}'' = e^{j\pi}\bs{s}^*$ and $\bs{s}''' = e^{j\frac{3\pi}{2}}\bs{s}^*$. Thus, we only need to optimize for a quarter of the total number of input vectors, resulting in $4^{M-1}$ times the algorithm is run per channel realization. 

Because of the still exponentially growing number the implementation has to be carefully considered. It would certainly be necessary to design custom hardware for the algorithms. Also we think parallel processing should be applied, since the process is easily parallelizable. With these and possible other techniques we are confident that the precoder can deliver good performance with short processing time, especially for a relatively small number of users.

In theory the required memory for the LUT increases exponentially with $M$. However, if $M$ is large only a fraction of possible input vectors $\bs{s}$ is sent within the coherence time. Therefore, it is sufficient to compute and store the vectors $\bs{x}$ only for the sent symbols.

We can summarize that theoretically the demands in processing time and memory increase exponentially. However, in a realistic implementation it is suggested that these demands can be substantially reduced. Therefore, the proposed method is a promising candidate for implementation in a real system.

\section{Conclusion}
\label{sec:conclusion}

We have presented a CE precoding technique that uses the SMSE criterion to map each input vector to an optimal transmit vector. We found that with different solution algorithms good performance with respect to BER can be achieved. When QPSK is chosen, the optimal precoder is the GPM RPC precoder. For 8-PSK the QGDM CEC precoder should be chosen. As a result we have a highly energy efficient system with low-resolution DACs and PAs can be operated in the saturation region. Through analysis of the number of iterations of the algorithms and the overall complexity we can conclude that the implementation should be carefully considered to achieve shortest possible processing time. Future work on this topic could include the employment of higher order modulation schemes, for example 16-QAM as well as the optimization for frequency selective channels. Also the robustness to channel estimation errors should be analyzed.

\bibliographystyle{IEEEtran}
\bibliography{refs}

\end{document}